High local genetic diversity and low outcrossing rate
in *Caenorhabditis elegans* natural populations

Antoine Barrière and Marie-Anne Félix*

Institut Jacques Monod, CNRS - Universities Paris 6 and 7,
Tour 43, 2 place Jussieu, 75251 Paris cedex 05, France

Tel: +33-1-44-27-40-88; Fax: +33-1-44-27-52-65
* corresponding author - e-mail: felix@ijm.jussieu.fr

**Summary**

**Background:** *Caenorhabditis elegans* is a major model system in biology, yet very little is known about its biology outside the laboratory. Especially, its unusual mode of reproduction with self-fertile hermaphrodites and facultative males raises the question of its frequency of outcrossing in natural populations.

**Results:** We describe the first analysis of *C. elegans* individuals sampled directly from natural populations. *C. elegans* is found predominantly in the dauer stage and with a very low frequency of males compared with hermaphrodites. While *C. elegans* was previously shown to display a low worldwide genetic diversity, we find by comparison a surprisingly high local genetic diversity of *C. elegans* populations; this local diversity is contributed in great part by immigration of new alleles rather than by mutation. Our results on heterozygote frequency, male frequency and linkage disequilibrium furthermore show that selfing is the predominant mode of reproduction in *C. elegans* natural populations, yet that infrequent outcrossing events occur, at a rate of approximately 1%.

**Conclusions:** Our results give a first insight in the biology of *C. elegans* in the natural populations. They demonstrate that local populations of *C. elegans* are genetically diverse and that a low frequency of outcrossing allows for the recombination of these locally diverse genotypes.

**Introduction**

Despite the status of the nematode *Caenorhabditis elegans* as a major biological model system, little is known about its biology in the wild, especially its reproductive mode and population structure. In the laboratory, *C. elegans* reproduces through self-fertile XX hermaphrodites and facultative XO males; males result either from rare X-chromosome non-disjunction during meiosis of the hermaphrodite germ line, or from outcrossing, which yields 50% male progenies. Sperm production occurs before oocyte production in hermaphrodites, and the number of sperm limits self-progeny number [1, 2]. A particularly long-standing question is whether males play a role in shaping the genetic structure and evolution of natural populations through sexual reproduction. Previous studies showed that males do not reproduce efficiently enough to be spontaneously maintained under laboratory conditions [3-5]. Selfing can strongly reduce genetic diversity in a species, because selection at one locus may drive to fixation the rest of the genome with it, due to strong linkage disequilibrium across the whole genome [6]. The overall known genetic diversity of *C. elegans* is low [7-13]. The question is thus raised whether local populations of *C. elegans* are genetically uniform and whether males and sexual reproduction occur. Understanding the biology of this model organism in its natural context is also a prerequisite to decipher many features of its development and behavior in the laboratory.

**Results and Discussion**
**Isolation of *C. elegans***

Although a worldwide set of single wild isolates of *C. elegans* has been collected over the last 40 years [14], this species has proven difficult to isolate and study in its natural environment. After extensive searches in a variety of habitats, we were able to find *C. elegans* in farmland and garden soil, compost heaps, and in association with diverse invertebrates. In contrast to other sampling procedures that isolate individuals after one to several generations in the laboratory ([15] and Sivasundar and Hey, personal communication), ours allowed us to directly isolate wild individuals within a few hours of substrate collection and thus to assess the stage, sex and genotype of each individual (Table 1).

*C. elegans* develops through a short embryonic stage followed by four juvenile stages (L1-L4). This life cycle takes 3.5 days at 20°C under laboratory conditions, but development can be much lengthened by an arrest in the L1 stage (in the absence of food) or in the dauer stage, a non-feeding alternative third juvenile stage, which in the laboratory is induced under conditions of low food, high crowding and/or high temperature. Dauer larvae are resistant to many stresses and can live for several months without feeding [16].

Predominantly we found *C. elegans* in the dauer stage, even in a putatively food-rich habitat such as compost heaps (where it occurred at densities up to 25 individuals per g; Table 1 and Fig. 1). We found some non-dauer stages (including non-dauer L3 larvae) in recent compost material, which indicates that *C. elegans* does feed and reproduce in compost heaps. Interestingly, several of these animals displayed features that are not normally seen in laboratory conditions, thus revealing ecological pressures that act on the species: starved or constipated animals, a transiently sterile adult, individuals that developed with a low number of intestinal cell divisions, individuals with undigested bacterial spores inside their intestine (not shown) or with bacteria inside the body itself, etc. (Fig. 1C).

*C. elegans* thus spends much of its time as a dauer. In rhabditid nematodes such as *C. elegans*, the dauer is a dispersal morph and may also associate with other invertebrates, parasitically, necromenically (feeding on the carcass of their host) or phoretically (dispersal) [17-21]. *C. elegans* has previously been found associated with diverse invertebrates [17]. We sampled a variety of macroscopic invertebrates and found *C. elegans* on snails (genera *Helix*, *Oxychilus*, and *Pomatias elegans*), isopods (*Oniscus asellus*) and a *Glomeris* myriapod (Fig. 1B). This suggests that, unlike some other nematodes, *C. elegans* does not have a narrow host specificity (Table 1). These associations are likely to contribute to the dispersal of *C. elegans* and may in addition be necromenic.

We found the close relative *Caenorhabditis briggsae* in the same type of habitat as *C. elegans*: farmland and garden soil, compost heaps, snails. In several locations, we even found *C. briggsae* to be co-occurring with *C. elegans* (whereas we never found other *Caenorhabditis* species such as *C. remanei*). *C. briggsae* and *C. elegans* may thus have similar habitats.

**Whole-genome polymorphism levels**

For genotype analysis, we collected one sample (10-50 cm$^3$ of substrate) from each of four locations (Merlet1, Le Blanc, Franconville, Hermanville) within France (Fig. 1A) and from each sample we isolated 11-12 individual *C. elegans* hermaphrodites, ensuring that they were derived from eggs produced prior to sample collection. At one location, we collected two more samples at distances of 10 to 30 meters apart (Merlet2-3). When analyzing the data, we thus distinguished six samples and four locations (Table 1). From each of these 55 sampled individuals, we established an isogenic strain by selfing. This is, to date, the only set of *C. elegans* strains derived from independent single individuals from the wild (other isolation procedures require several generations before individuals are sampled; [15] and Sivasundar and Hey, personal communication).

Previous DNA sequence and microsatellite studies on strains sampled worldwide found a low overall level of polymorphism in *C. elegans* and similar sets of genotypes in several world regions (e.g. England, California, Australia) [7-12]. The divergent CB4856 (Hawaii) strain was estimated to differ from the reference strain N2 (England) at an average of 1/840 to 1/370 nucleotide sites of nuclear DNA [8, 11, 22]. This within-species diversity is low compared to that found in a dioecious outcrossing species of the same genus, *C. remanei* [9], and to other selfing soil nematode species (*Pristionchus pacificus*, AFLP and sequencing data [23, 24]; *Oscheius tipulae*, D. Baïlle and M.-A. F., unpublished AFLP data).

In order to measure genetic diversity between our sampled individuals, we used Amplified Fragment Length Polymorphism (AFLP), a sensitive method that detects low levels of polymorphisms in a random genome-wide manner [25] (Table S1). Because AFLP does not provide co-dominant markers, we analyzed haploid sets of genomes present in the diploid form in the isogenic strains and did not try to record heterozygous loci with this method. We first analyze below the genetic diversity over the whole dataset and relative to CB4856, and then analyze the level of local genetic diversity within each sampling location.

We found a low overall genetic diversity in our samples, with only 31/149 (21%) AFLP fragments being polymorphic over the whole dataset and an overall Nei diversity of AFLP profiles $Hj = 0.049$ (i.e. an average divergence of 4.9% of the fragments between any two genomes sampled at random in the dataset; Table 2). For comparison, two divergent strains of *Pristionchus pacificus* displayed an AFLP polymorphism level of 45% [23, 24]. As references, we included the N2 and CB4856 strains in the AFLP analysis and found a polymorphism at 7/149 (4.7%) of the AFLP loci; each of the corresponding 14 alleles was found in our dataset over the four sampling locations in France (Table S1), which thus presumably cover a significant part of the diversity found worldwide, as was previously observed for samples from England or California [7-12]. The AFLP diversity that we find corresponds approximately to a nucleotide diversity of $0.8 \times 10^{-3}$, which is compatible with previous sequence data - which so far relied on very few polymorphisms [9, 10, 13]. This value is approximately 20 and 10 times lower than those of *Drosophila melanogaster* [26] and of the partial selfer *Arabidopsis thaliana* [27], respectively, and similar to that of humans [28]. Our data thus confirm, at the scale of a few hundred kilometers over our whole dataset, the low level of global genetic diversity in *C. elegans*.

Given this low global diversity of *C. elegans*, and relative to it, the level of local genetic diversity was surprisingly high at the scale of a few centimeters (a sample) or meters (a location) (Table 2). Out of the four samples of 11-12 individuals, two (Franconville and Hermanville) displayed many polymorphisms. The two other samples (Le Blanc and Merlet1) displayed fewer polymorphisms, three out of four being singleton polymorphisms that may have originated from recent mutations. We also found a high genetic diversity between the three samples within the Merlet location.

How does this local diversity compare to the overall genetic diversity within *C. elegans*? Strikingly, except for the Blanc location, the within-location diversity estimates are comparable to the estimated genetic distance between N2 and CB4856 (7 loci): there are on average 3.9 and 3.5 differences between pairs of strains within the Hermanville and Franconville samples, respectively, and 7.2 differences within the Merlet location (taking into account the three samples). Moreover, each of the 7 loci that differ between N2 and CB4856 is also polymorphic within at least one of the locations (Table S1). Local populations of *C. elegans* thus present a remarkably high level of polymorphism when compared to the diversity found worldwide.

Sequencing of mitochondrial DNA on a worldwide set of *C. elegans* strains previously distinguished two well-resolved clades, called mitochondrial clades I and II, which represent an ancient divergence in mitochondrial lineages [11]. To monitor the mitochondrial genotype of our sampled individuals, we chose 3 clade-specific polymorphisms that could be tested on the same PCR product by restriction digestion. We find that both types of mitochondrial genotypes co-occurred in three out of four sampling locations (Table 2). A significant part of worldwide diversity of *C. elegans* is thus found within a single garden or compost heap.

**Role of migration**

The high local molecular diversity that we observe in several *C. elegans* populations may have arisen through local mutation or by migration of divergent individuals (either at the origin of a population or as a subsequent import of genotypes). Of the 31 polymorphic loci, we found 18 to be specific to particular sampling locations, including 7 singletons (Table 2); these polymorphisms may be the result of local mutation. On the other hand, 12 polymorphisms, plus those in the maternally-transmitted mitochondrial DNA, are shared among different sampling sites. When considering the origin of genetic diversity within one location (Table 2), more than half of its local polymorphisms are shared with another location: for example, 6/11 polymorphisms that are found within the Franconville sample are shared polymorphisms (i.e. both alleles are found outside the Franconville sample; Table 2).

Migration appears however insufficient to homogenize allele frequencies at a given time over different sampling locations. Indeed, a large proportion of the genetic variation between individuals was found between the four sampling locations (expressed as a proportion of the total variance, taking the largest of the three Merlet samples: $Fst$ = 0.78; p<0.001; using all Merlet individuals: $Fst$ = 0.83; p<0.001), indicating that the genotype pools were highly distinct between the samples. Also, none of the 24 different multi-locus genotypes (combinations of alleles at the different genetic loci) that we found was shared between sampling locations (with the exception of the strain JU438, sampled at a later time in Hermanville, which was identical to JU298 from Le Blanc), though each was often found several times within a sampling location (Table 2). Since selfing following population bottlenecks reduces intra-sample diversity and thereby results in a high $Fst$ value [29], we also tested for spatial structure between sampling sites by pooling all identical multilocus genotypes. Even in this case, similar alleles are found much more often within a site than at random ($Fst$ = 0.44; p = 0.004). Taking into account the presence of shared alleles rather than their frequency (thus reducing this influence of recent selfing on allele frequencies), we also find a highly significant differentiation between the four locations (p<0.01 for all pairwise tests) [30]. Thus, although migration is important in fostering local diversity, it does not appear sufficient to prevent population structure at the spatio-temporal scale our sampling considers (over 100 km between locations, one time point).

Strikingly, very distinct sets of multilocus genotypes were found in the three samples from the Merlet location, even though they were collected within a few meters of one another, indicating strong structure at this scale as well ($Fst$ between the three Merlet samples = 0.96; p<0.001). The observation of a very strong structure at a small scale within Merlet, similar to that observed at a much larger scale, suggests that demographic bottlenecks in this selfing species may be a sufficient explanation of the structuring between sampling sites. Moreover, we found no correlation between geographical distance and Nei genetic distance in the six samples, (which cover a scale of ten meters to hundreds of kilometers; Mantel test, p = 0.94), thus further suggesting that the migration rate is large enough that isolation by distance may not contribute substantially to the structuring at the larger scale. Much of this structure may be the result of recent population demographic bottlenecks or an indirect consequence of selfing (resulting in a lower effective population size, especially following selective sweeps) [6].

Similar sets of several divergent genotypes were previously found in Australia, America and Europe, with no correlation between genetic and geographic distances [7, 11, 12], suggesting that *C. elegans* can migrate at a long range. The finding of *C. elegans* in human-associated habitats makes it possible that human activity is a major factor in present-day *C. elegans* migration patterns.

In summary, our observations of a large proportion of shared polymorphisms between locations yet a strong apparent spatial structure are thus not incompatible [29, 31, 32]. The sharing of polymorphisms between locations indicates a continuous input of migration and/or a large population size at population foundation with sharing of ancestral polymorphisms. The low diversity levels of some of the samples may reflect a lack of demographic stability in some populations, with transient bottlenecks that reduce genetic diversity [33].

**Genomic reassortments and outcrossing**

In addition to overall genome diversity, our AFLP data provide allele combinations at the different loci for each sampled haploid genome. These multi-locus genotypes depend on the shuffling of allele combinations over loci (spanning the six chromosomes), and thus on the frequency of cross-fertilization in this partially selfing species. Within each sampling location, combinations of alleles are clearly non-randomly distributed, given local allele frequencies, and are in strong linkage disequilibrium (Tables 3 and S1). At a larger scale over the four locations, taking into account each multilocus genotype only once, the linkage disequilibrium is also significant (Table 3), indicating limited outcrossing. Since even infrequent outcrossing would rapidly break down linkage disequilibrium between markers (except tightly linked ones) [29], these results strongly suggest that selfing is the most frequent reproductive mode in these *C. elegans* populations.

In *C. elegans*, outcrossing occurs only when hermaphrodites are inseminated by males, which result either from rare events of X-chromosome meiotic non-disjunction (at frequencies in the 0.03-0.3% range for wild strains in laboratory conditions [14]), or from outcrossing (50% of the cross-progeny is male). As males are both a prerequisite for, and the consequence of, outcrossing, we assessed whether outcrossing may occur by estimating the proportion of males in natural populations. In all sampled populations, males occur at a very low frequency that is compatible with the spontaneous X-chromosome non-disjunction rate (Table 1). Moreover, none of the six adults that we collected (Table 1) gave rise to a large proportion of males in its progeny, suggesting that they had not mated with males in the wild. Thus, in addition to the strong linkage disequilibrium, this low frequency of males further indicates that selfing is the overwhelmingly predominant mode of reproduction in these natural *C. elegans* populations.

Is there, on the other hand, any positive evidence for cross-fertilization by males? For a set of strains sampled worldwide, Denver et al. found that phylogenetic trees constructed using mitochondrial versus nuclear DNA sequence data were largely congruent, suggesting that recombination between divergent *C. elegans* genomes is exceptional at this scale [11]. However, our data do indicate reassortment events between the polymorphic loci, at least at the between-location level, as anecdotally noted by others [7-12, 15]. Firstly, our results, unlike those of Denver et al. [11], show no global concordance between nuclear and mitochondrial polymorphisms: a tree based on the distances between multi-locus AFLP genotypes shows no clustering of mitochondrial genotypes as shown in Denver et al. [11]. The separation of the 24 AFLP multilocus genotypes according to their mitochondrial genotype results in a significant structure ($Fst$ = 0.19, p<0.001), which suggests that the linkage between mitochondrial and nuclear loci is not fully broken within our sample. However, this may also be due to sample structure. Even among the polymorphisms studied by Denver et al., we found discordance, including 7 Non-Plugger strains with clade-II-like mitochondrial markers (Table S1). Secondly, many reassortments between polymorphic loci can be detected by the occurrence of all four combinations of genotypes at two loci (4-gamete test, [34]; Table S2). Out of the 25 non-singleton polymorphisms (including the mitochondrial genotype), 23 show evidence of reassortment with at least one other locus in the whole dataset. We were furthermore able to identify putative recombination events within Merlet and Franconville, on polymorphisms that are apparently specific of each of these locations. None of the putatively 'new' alleles of location-specific polymorphisms are shared by worms with different mitochondrial genotypes; given that the separation of the two mitochondrial clades probably pre-dates many of the sampling-location specific polymorphisms, this strengthens the finding that genome reassortments can mostly be detected at the supra-location level (Table S2). Using the published *C. elegans* genomic sequence, we could identify most AFLP fragments present in the N2 genome (123/133 of the theoretical fragments including 9/10 of the polymorphic ones; Table S1). One pair of loci on chromosome V (*mfP4* and *mfP8*, 10.51 cM apart) and 2 pairs on chromosome IV (*mfP2* and *mfP10*, 39.69 cM, and *mfP6* and *mfP10*, 42.24 cM) show evidence of recombination in the 4-gamete test. Thus, recombination events are detectable between and within chromosomes. Even though it is infrequent, outcrossing with males thus occurs in *C. elegans* natural populations.

From our linkage disequilibrium data, we estimated an outcrossing rate, using a model of population of effective size $N_e$ (estimating $N_e$ under a model of mutation-drift in Table 2). We calculated the mean linkage disequilibrium between pairs of loci, (for Hermanville, mean($r^2$)=0.2906, 95% confidence interval: 0.2489-0.6881; for Franconville, mean($r^2$)=0.5175, 95% confidence interval: 0.1667-1). Using a theoretical recombination rate $c_{th}$ calculated as 0.44, we estimated the outcrossing rate to be one outcrossing every 7,351 generations for Hermanville (95% confidence interval: 5,947-19,601 generations) and every 17,365 generations for Franconville (95% c.i.: one every 3,239 generations to complete selfing) [31, 35].

We directly assessed heterozygote frequency in freshly isolated wild populations using microsatellite repeat length polymorphisms (locus II-R and IV-R in ref. [15]). Three out of the four screened populations were found to be polymorphic for one or both loci, with a strong differentiation between the populations, even at a 1 km scale between Primel and Sainte-Barbe, thus confirming the conclusions drawn from the AFLP data obtained on other populations (Table 4). In two distinct populations, heterozygotes could be found, proving the occurrence of outcrossing in the wild. At least for the II-R heterozygotes, the possibility of them arising through recent mutation is unlikely because both alleles were also found in at least one other location (see also [15]). From the observed microsatellite diversity, we calculated the inbreeding coefficient, from which we estimated the inbreeding and outcrossing rates [36]. Selfing rates for the three populations are consistent, ranging from 1 to 0.98 (Table 4). From the global inbreeding coefficient, we calculated the mean selfing rate over the four populations and found $S_{mean}$ to be 0.987, or 1.3% of outcrossing.

Our outcrossing rate estimates using linkage disequilibrium are lower by one order of magnitude than those determined by measuring heterozygote frequencies, but it must be noted that both methods are not equivalent: while the outcrossing rate calculated from heterozygote frequency is a measure at one timepoint, the estimate based on linkage disequilibrium is a measure over a longer time period, taking into account population structure, linkage with loci under selection (which should actually result in an underestimation) and changes in outcrossing rate over time. However, the estimate based on linkage disequilibrium is less direct and relies on several assumptions and estimations, weakening the confidence in the conclusion.

It is possible that outcrossing may differ depending on ecological conditions; for example, the frequency of X-chromosome non-disjunction is increased at high temperatures in the laboratory [2]. While we cannot generalize for all populations (especially in other ecological conditions), our results show that at least some natural *C. elegans* populations are maintained with a high selfing rate. The selfing mode of reproduction of *C. elegans* makes it a potentially good colonizer: in a selfing species, a single individual is required to found a population. A selfing population is subsequently less likely to be subject to inbreeding depression after a bottleneck, because strongly deleterious recessive mutations are previously eliminated more rapidly than in a dioic species [37]. The spatial distribution of genotypes that we describe is consistent with a relatively high rate of migration together with local demographic bottlenecks, and thus with high metapopulation dynamics of extinction-recolonization. That *C. elegans* is a colonizer is also compatible with its finding in discontinuous temporary habitats and its frequent occurrence in a resistant dispersal form.

**Conclusions**

We describe the first extensive sampling of *C. elegans* individuals directly from natural populations, which allows us to determine biological features of wild *C. elegans* individuals and the genetic structure of *C. elegans* populations. We mostly find *C. elegans* in the dauer stage, in compost heaps and on a variety of carrier invertebrates (snails, isopods, etc.). We find, at least in some populations, a high genetic diversity (compared with the low global genetic diversity in the species) at the spatial scale that is relevant for mating. Our results suggest that *C. elegans* mostly reproduce by selfing in these populations, yet that infrequent outcrossing occurs. The rare males in these populations mate rarely, but if so, will often mate with a hermaphrodite of a different genotype. These rare outcrossing events may thus recombine allelic variants at multiple loci (thus participating in the elimination of weakly deleterious mutations) and contribute to the evolution of the *C. elegans* genome.

**Experimental procedures**
**Sampling and identification of *C. elegans***

The samples were collected in: Franconville, Val d'Oise, 48.98°N, 2.23°E (a Parisian suburb garden); Le Blanc, Indre, 46.63°N, 1.07°E (a garden in a small town/village); Hermanville, Calvados, 49.28°N, 0.32°W (idem); Merlet, hamlet near Lagorce, Ardèche, 44.45°N, 4.42°E (the garden of an isolated farmhouse); Le Perreux-sur-Marne, Val-de-Marne, 48.85°N, 2.50°E (a Parisian suburb garden); Primel-Trégastel and Sainte-Barbe, 48.8°N, 3.48°W (hamlets near Plougasnou, Finistère; 1 km from each other) (Figure 1). In Merlet, samples were collected from different locations in the garden, Merlet 1 and Merlet 2 being distant of 15 meters, Merlet 3 a further 12 meters away from Merlet 2. During our extensive hunts, we also found *C. elegans* in one more compost heap near Paris, in leaf litter in a garden in Frechendets, 43.07°N, 0.25°E (Hautes-Pyrénées), in garden soil and snails in Le Blanc and Hermanville, in snails in Franconville and in soil of an orchard in Madeira (yielding strain JU258). Despite collecting hundreds of samples worldwide over several years, we failed to find it in any less humanized habitat. At least in the world regions that were most extensively sampled, *C. elegans* thus appears to have become a commensal of human activity.

Soil and compost were collected in 10-50 ml tubes. Samples (0.5-2 g) were placed around an OP50 *E. coli* lawn on NGM plates (55 or 100 mm diameter) [16], and water was spread on them (ca. 1 ml per g). Snails were washed in water to remove adhering soil/tree bark, crushed and placed onto NGM plates. Smaller invertebrates such as isopods were sacrificed by cutting them in half. Nematodes were picked as they crawled out of the substrate in the first three days after sampling. Most (about two-thirds) *C. elegans* individuals were recovered within 1 min to 10 hrs (Table 1B). All stages could be recovered in a quantitative manner, except for the embryonic stages, which would only be subsequently recovered as L1 larvae.

A first genus-level determination was conducted with a dissecting microscope and a Nomarski microscope using several criteria: morphology of buccal cavity, pharynx (a strong and round middle bulb), gut (light-colored gut, with large cell nuclei), vulva lineage, tail (elongated) [38]. Presumptive *Caenorhabditis* individuals were then bred on NGM plates. Isogenic strains were established by selfing of hermaphrodites.

Species identity was tested by two criteria, i) a mating test: 3-5 virgin hermaphrodites (descendants of the isolated wild individual) were placed with 3-5 males of the N2 *C. elegans* reference strain (or of a GFP-labeled strain derived from N2); successful mating was scored by the occurrence of a large proportion of males in the progeny; ii) a species-specific PCR test: Species-specific PCR primers were designed in the gene *glp-1* [9] so as to amplify a different fragment size in *C. elegans*, *C. briggsae* and *C. remanei*. *C. elegans*: ccgcttggattctatggattg and ctctccttgctccgggattg, product: 208 bp. *C. briggsae*: gaacctgcgagtgcatgtac and ccgtctgcaaacgaacgggc, product: 302 bp. *C. remanei*: caacggaggtatctgctcag and ccgccgtcaaatttgcattc, product: 391 bp. PCR was performed using all 6 primers in the same reaction. All *C. elegans* strains in Table 1A were tested using both mating and PCR tests. Among the animals in Table 1B, we did not test every single *Caenorhabditis* individual: for example, we performed a mating test on a subsample of 12 animals in the sample "Le Perreux 5 Oct 04", 12 (compost) + 4 (snails) in "Franconville 6 Oct" 04, 12 in "Sainte-Barbe", 12 (compost) + 22 (isopods) in "Primel", etc., and all were *C. elegans*. *C. briggsae* was never found in these locations. It is thus highly probable that all other hermaphroditic animals that we screened as *Caenorhabditis* were *C. elegans*. It is striking that *C. elegans* is found almost exclusively in the dauer stage even in samples where all stages of other nematode species of the same family are found.

*C. elegans* and *C. briggsae* were found co-occurring in Merlet, Hermanville and several other locations, but we never found any other *Caenorhabditis* species. We found *C. briggsae* in soil, compost heaps and snails: a decomposing *Helix aspersa* in the Garden of the Natural History Museum, Paris, and a *Oxychilus* sp. in Merlet 2.

Strains are available on request (www2.ijm.jussieu.fr/worms).

**AFLP analysis**

Worms from recently starved OP50 culture plates were rinsed several times for several hours in a large volume of M9 to avoid contamination from bacterial DNA. Genomic DNA was prepared with a Qiagen DNeasy Tissue kit, digested with EcoRI and Tru1I (a MseI isoschizomer) and ligated with specific adapters [25]. The ligation was diluted 1:10 and amplified by PCR with primers specific to the adapters. A second amplification round was carried with fluorescent primers corresponding to the adapters plus 5 selective nucleotides. AFLP fragments were run on an ABI 3100-Avant Prism capillary sequencer. All reactions were run in duplicates from the DNA digestion on. AFLP profiles were scored with Genographer v.1.6.0 (size range 50-400 bp). For analysis, we only kept fragments that amplified reproducibly in duplicates of each strain.

To avoid counting a polymorphism twice, we used AFLP primer combinations distant by more than one nucleotide, and when we noticed complementary profiles of presence/absence of fragment with the same primers due to a small insertion-deletion (indel), we included only one of them (4 instances).

In order to convert the AFLP diversity ($Hj$) into nucleotide diversity ($p$), we assumed as an approximation that an AFLP polymorphism can be the result of a substitution or indel in the restriction sites plus selective nucleotides (15 nucleotides) (presence/absence of a fragment) or to an indel in the fragment itself (change in size, including shifts outside the scored range). The ratio of the two events in *C. elegans* was estimated to 3.9 substitutions per indel in ref. [11] (50 kbp sequence in several wild strains) and 3.0 in ref. [8] (5.4 Mbp in CB4856). The average size of our AFLP loci being 176 bp, we expect between 176/(15x3.9) and 176/(15x3.0) = 3.0 and 3.9 indel polymorphisms, respectively, for each point mutation polymorphism (we only found 4/31 instances of fragment size shift, probably because some result in a superposition with another fragment or in a shift of the fragment outside the scored range). We thus calculated the average pairwise nucleotide diversity by dividing the AFLP diversity by 15 (selective nucleotides) and applying the correction for indels. Because the diversity level in our sample is low (and because of the unavoidable approximations due to the correction for indels), we did not correct for fragment size homoplasy as in ref. [39-41]. As an internal control, we checked that our AFLP distance between the strains N2 and CB4856 (7 polymorphic loci) was compatible with the known nucleotide distance between these strains.

Nei's diversity for restriction-site polymorphisms ($Hj$) and $Fst$ values were calculated from AFLP allele frequencies after [42] with AFLPsurv v.1.0 (Vekemans, X. 2002, Université Libre de Bruxelles, Belgium).

The AFLP fragments of the N2 genome were identified using the e-PCR tool in Wormbase (release WS132, available online at www.wormbase.org using, as primers, the restriction sites and selective nucleotides used in the AFLP analysis. For each positive result, we checked for the presence in the fragment of an EcoRI or MseI restriction site. Fragments showing no restriction site, and therefore constituting a virtual N2 AFLP profile, were scored against the experimental N2 AFLP profile. Out of 133 theoretical fragments in the 71-395 bp range, 123 (92%) unambiguously matched observed fragments.

**Mitochondrial genotype**

A region of mitochondrial DNA was amplified by PCR with the primers aaataagtatgtttcttttcgcag and attttgattttcttacgataccnc and digested by HinfI (3 or 1 cuts respectively) or ApaI (1 or 0 cuts respectively). Digestion products were run on a 0.8% agarose gel.

**Plugging test**

The Plugging polymorphism was assayed as in ref. [14]. Cultures containing males were produced in each strain (after heat-shock if required) and maintained by crossing. Two hermaphrodites and three males of the tested strain were placed on the same culture plate for 24 hours; the presence of a mating plug on the hermaphrodite vulva was then scored under the Nomarski microscope. In the absence of a mating plug, the occurrence of mating was checked by the presence of a high proportion of males in the cross-progeny.

**Analysis of linkage disequilibrium**

Pairwise linkage disequilibrium was calculated in Excel using R [43]. Multilocus linkage disequilibrium was calculated with Lian 3.1 [44]. Tests of the presence of all four combinations of alleles at 2 loci (4-gamete tests) were performed under Excel, including N2 and CB4856 in the dataset. We assumed that the absence of a band was homologous, which is a reasonable approximation, given the low sample-wide AFLP diversity.

**Estimation of the outcrossing rate from linkage disequilibrium**

We estimated the outcrossing rate using the correlation between pairs of loci. Linkage disequilibrium is linked to $c$, the recombination rate, by the equation $E(r^2)=1/(1+4N_e c)$, $r^2$ being the squared correlation coefficient between two loci [35]. The observed recombination rate, $c_{obs}$, was estimated by taking $E(r^2)$ as the mean of $r^2$ values calculated over all pairs of polymorphic loci (excluding polymorphisms occurring at frequencies under 10%) and taking $N_e$ estimates from Table 2. The outcrossing coefficient $(1-F)$ is linked to the observed recombination rate and the expected recombination rate, here called $c$ theoretical ($c_{th}$), by $c_{obs} = c_{th}(1-F)$ [31]. $c_{th}$ was estimated by simulation, considering the physical length of chromosomes and assuming a constant recombination rate. The selfing rate $S$ was calculated from $S=2F/(1+F)$, and the outcrossing rate as $1-S$. To calculate a confidence interval, we created a theoretical population with the genotype frequencies observed in our sample, then took 1,000 samples of 12 individuals and calculated the mean squared correlation coefficient. From this distribution, we took a 95% confidence interval.

**Microsatellite amplification**

We isolated single wild individuals and let them reproduce over two generations on a standard 55 mm culture plate, then washed the plates with M9 and used 1/15[th] of the worms for DNA extraction, while the rest was frozen as stock. DNA was extracted using a guanidine isothiocyanate precipitation on silica, and resuspended in 100 µL $H_2O$. Amplification was carried with Eppendorf Taq polymerase in 10 µL volumes, in two steps with the following cycle: first step at 2 min of initial denaturation at 92°C, then 40 cycles with 20 sec at 92°C, 30 sec at 60°C, 30 sec at 72°C, then 10 min of final extension at 72°C. Primers used were as described in ref. [15], with an M13 forward tail added to the forward primer. The second step was carried at an annealing temperature of 52°C, and the locus-specific forward primer replaced by a M13 forward primer labeled with the Hex or 6-Fam fluorochromes. Electrophoresis was then performed on an ABI 3100-Avant Prism capillary sequencer with molecular weight markers. Possible heterozygotes at the II-R locus were then confirmed by thawing frozen stocks and genotyping individual progeny; in each case, we found homozygotes for both alleles and heterozygotes in the progeny. Six of the microsatellites described in ref. [15] were initially used, but II-R and IV-R were selected because they amplified reproducibly and were polymorphic enough for an estimation of outcrossing frequency.

**Estimation of the outcrossing rate from heterozygote frequency**

Computation was done with Fstat [36]. We used the $f$ estimation of $Fis$ after [45]. Gene diversity $He$ was calculated with Arlequin [46]. The selfing rate $S$ was calculated from $S=2Fis/(1+Fis)$, and the outcrossing rate as $1-S$.

**Acknowledgements**. We thank the many people who helped us with sampling, especially C. Pieau, B. Faverjon and A. Borély, F.-X. and O. Barrière, M., B., C., I. and E. Félix, F. Back and J.-A. Lepesant. We are very grateful to M. Veuille, D. Higuet, B. Charlesworth, C. Bazin and C. Braendle for helpful discussions and key suggestions for data analysis. We thank A. Sivasundar and J. Hey for communication of unpublished results. We also thank D. Charlesworth, A. Cutter, E. Dolgin and H. Schulenburg for very helpful comments on the manuscript. This work was supported by the CNRS and the Ministry of Research of France through a predoctoral fellowship to A.B. and a 'Biological Resource Center' grant.

**Figure 1. Sampling of *C. elegans* populations.** A. Distribution of sampled sites in France. The locations where samples for AFLP analysis were collected are in bold letter. B. Invertebrate species from which *C. elegans* was collected. Top: snails; from left to right: *Helix*, *Pomatias* and *Oxychilus* species (scale bar: 1cm). Bottom: *Oniscus asellus* isopod (scale bar: 1 mm). C. Nomarski micrographs of *C. elegans* individuals at the time of isolation. From top to bottom and left to right: two dauer larvae with characteristic alae on the cuticle (left) and male-specific cell division patterns in the rectal region (right; one of the two males in Table 1); a starved L4 larva, with few intestinal storage granules and few intestinal cell nuclei; an adult with abnormal eggs (after feeding on *E. coli* for several hours, it laid about 65 developing embryos during the next day), a L4 larva with internal bacteria, and a constipated L4 larva.

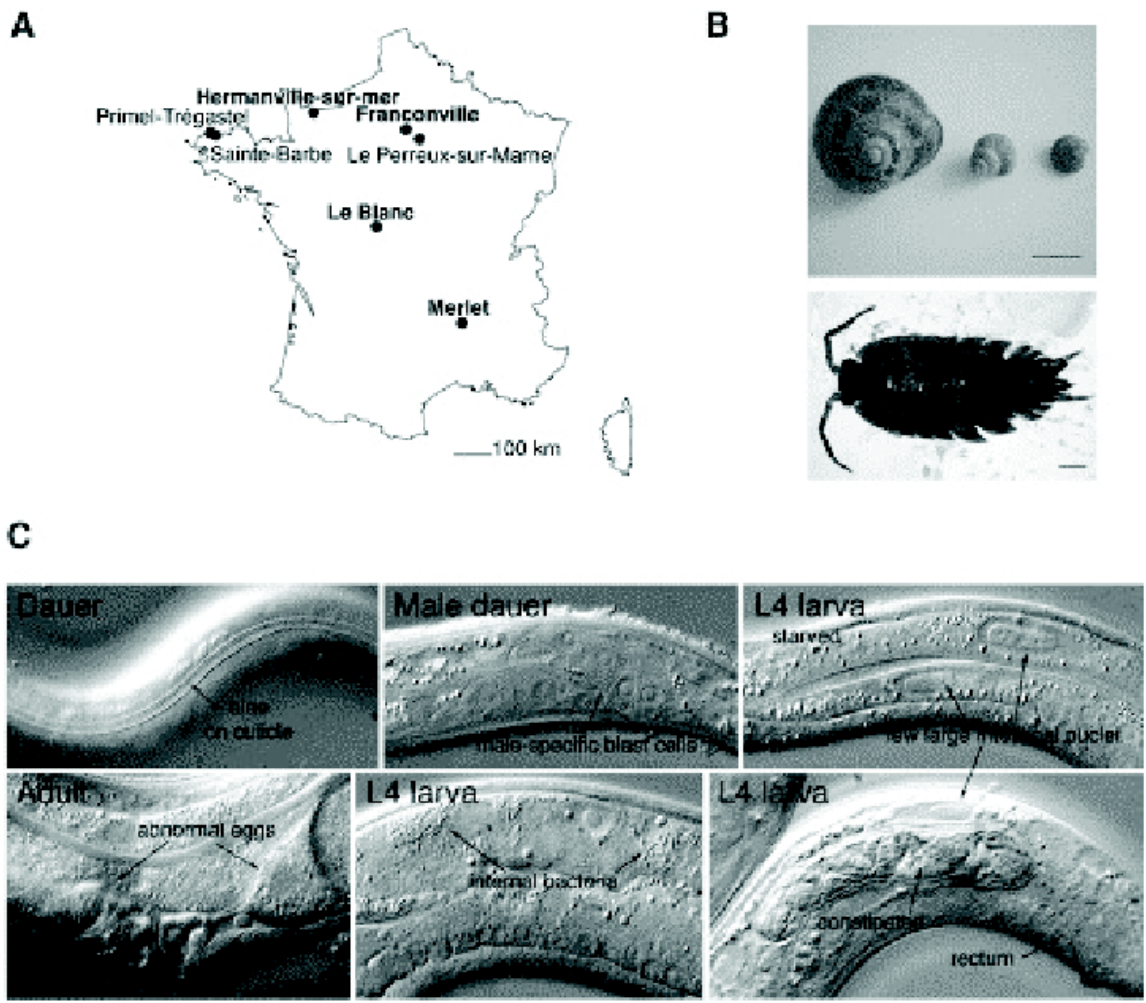

**Table 1. Habitat, sex, developmental stage and density of *C. elegans* individuals isolated from the wild.** A. Samples used in the AFLP analysis. B. Sex, developmental stage and density. In the substrate column, the numbers after the invertebrate refer to the number of animals carrying *C. elegans* dauers (for example 4/9 snails). 'Sex' and 'Developmental stage': are only included worms that were seen on the day of plating (the day of collection, or the next day for the two bottom lines). 'Total': total number of *C. elegans* in the sample after exhaustive examination for 3 days. 'H': hermaphrodites/total); 'M': males/total; 'D': dauer stage/total; '-': we did not determine the corresponding number. *: One of the two males sired progeny when placed with *unc-119(ed3) C. elegans* hermaphrodites; the other one is shown in Fig. 1B. #: this compost sample contained less than one *C. elegans* per 100 nematodes, whereas only one non-*C. elegans* nematode individual was found on the isopods.

| A. Samples used in the AFLP analysis | | |
|---|---|---|
| **Location** | **Date** | **Substrate (corresponding isogenic strains)** |
| Franconville | 16 Sep 02 | compost (JU360-371) |
| Hermanville | 22 Sep 02 | compost (JU393-402,406-407) |
| Le Blanc | 25 Aug 02 | compost (JU298-310) |
| Merlet 1 | 8 Sep 02 | soil (JU318-321); snails (JU311-317) |
| Merlet 2 | 8 Sep 02 | *Helix* snails on a mulberry tree (JU322,323,342,347) |
| Merlet 3 | 8 Sep 02 | *Glomeris* milliped in compost (JU343-346) |

| B. Sex, developmental stage and density | | | | | |
|---|---|---|---|---|---|
| **Location** | **Date** | **Substrate** | **isolated within: 10 hrs** | | **3 days** |
| | | | Sex | Dev. stage | Total |
| Franconville | 20 Jul 04 | compost | H: 3/3 | D: 3/3 | - |
| Franconville | 6 Oct 04 | compost 13 g | H: 81/81 | D: 81/81 | 130 |
| | | snails (4/9) | - | - | - |
| Franconville | 20 Oct 04 | compost 10 g (top of heap) | H: 65/65 | D: 63/65 1L1, 1 L4 | 99 |
| Franconville | 28 Oct 04 | compost 33 g (top of heap, recent material added) | H: 748/750 M: 2/750* | D: 717/750 10 L1, 4L1/L2, 6 L2, 9 L3, 3 L4, 1 adult | - |
| Franconville | 2 Nov 04 | compost 17 g (top of heap) | H: 104/104 | D: 78/104 8 L1, 6 L2, 2 L3, 5 L4, 5 adults | 89 in 9 g |
| Le Perreux | 7 Jul 04 | compost | H: 6/6 | D: 6/6 | - |
| Le Perreux | 5 Oct 04 | compost 17 g | H: 28/28 | D: 28/28 | 36 |
| Primel | 3 Oct 04 | compost ca.15 g | - | - | 32# |
| | | isopods (7/7) | H: 13/13 | D: 13/13 | 17 |
| Ste-Barbe | 3 Oct 04 | compost ca.15 g | - | - | 98 |
| | | isopod | - | - | 1 |

**Table 2. Local populations of *C. elegans* are polymorphic.** We distinguish four sampling locations and six samples; results for the Merlet sub-samples 1 to 3 are indicated below those obtained by pooling them (only Merlet1 has a sample size that is comparable to the samples in the three other locations). All: all 55 strains pooled before analysis. Nei's AFLP diversity is defined as the average proportion of loci that differ between pairs of strains. Nucleotide diversity is estimated using a correction for indels, assuming a ratio of 3.9 single nucleotide substitutions to 1 indel [11]. Mitochondrial genotypes scored as I* correspond to a loss of the *ApaI* site, compared with the previously characterized clade I sequence [11]. Singletons refer to genotypes found only once in the dataset, while shared polymorphisms refer to polymorphisms found within several populations. Location-specific polymorphisms are those found only within a single location (including data for N2 and CB4856). Effective population sizes are calculated as $Ne = \pi/4\mu$ [47], using the estimated nucleotide mutation rate $\mu = 2.1 \cdot 10^{-8}$ per nucleotide per generation for *C. elegans* [48] (note that this effective population size is that of a fictitious population that would display the corresponding diversity under mutation-drift equilibrium). We find a range that is compatible with independent estimates using microsatellite data (Sivasundar and Hey, personal communication).

| Location | Franconville F | Hermanville H | Le Blanc B | Merlet (total) | | | All |
|---|---|---|---|---|---|---|---|
| Sample | | | | M1 | M2 | M3 | |
| Strains | 12 | 12 | 12 | 19 | | | 55 |
| | | | | 11 | 4 | 4 | |
| Polymorphic bands (total scored : 149) | 11 | 11 | 1 | 19 | | | 31 |
| | | | | 3 | 2 | 0 | |
| Nei diversity of AFLP loci $Hj$ expressed in % (standard error) | 2.4 (0.8) | 2.6 (0.7) | 0.1 (0.1) | 4.8 (1.1) | | | 4.9 (0.9) |
| | | | | 0.3 (0.2) | 0.6 (0.4) | 0.0 (0.0) | |
| Nucleotide diversity ($\pi \times 10^3$) | 0.39 | 0.43 | 0.02 | 0.80 | | | 0.81 |
| | | | | 0.06 | 0.10 | 0.00 | |
| Effective population size ($N_e$) | 4,600 | 5,100 | 200 | 9,500 | | | 9,600 |
| | | | | 700 | 1,200 | (1) | |
| Multi-locus genotypes | 7 | 7 | 2 | 8 | | | 24 |
| | | | | 4 | 3 | 1 | |
| Mitochondrial genotype markers | I + II | I + II | II | I + I* + II | | | I + I* + II |
| | | | | I | I* | II | |
| Singletons | 2 | 2 | 1 | 2 | | | 7 |
| | | | | 2 | 0 | 0 | |
| Location-specific polym. | 5 | 5 | 1 | 7 | | | 18 |
| | | | | 2 | 0 | 0 | |
| Shared polymorphisms | 6 | 6 | 0 | 12 | | | 13 |
| | | | | 1 | 2 | 0 | |

**Table 3. *C. elegans* populations display a high linkage disequilibrium.** For each sampling location (except Le Blanc), linkage disequilibrium tests between AFLP loci (using in addition the mitochondrial genotype as one locus) were performed using all individuals (n: number of genotypes). Because the linkage disequilibrium within sampling locations is very strong, we also tested for linkage disequilibrium at a larger scale, across the four locations; for this analysis (three bottom lines), we counted each multi-locus genotype once only (i.e. the most conservative test). Left part, pairwise tests: linkage disequilibrium was tested for each pair of loci; to account for false positives due to the high number of pairwise comparisons, we show in brackets the number of comparisons that are expected to be significant at 1% in a random dataset (which is much lower than the observed value in each case, thus demonstrating significant linkage disequilibrium in the data), and the number that remain significant at 5% after a Bonferroni correction (another method that accounts for multiple testing). Each *C. elegans* chromosome has a genetic length of 50 centiMorgans and only a few pairs of AFLP loci would be expected to show linkage disequilibrium if outcrossing occurs. Among the 9 mapped polymorphisms, linkage disequilibrium tests were negative for 6/8 intra-chromosomal pairs, and positive for the closest pairs *mfP2* vs. *mfP5* (genetic distance = 0.31 cM; p = 0.01 after Bonferroni correction) and *mfP2* vs. *mfP6* (2.55 cM; p = 0.004) on chromosome IV. Of all polymorphic loci, only one shows linkage disequilirium with the mitochondrial genotype: *mfP9* in the Hermanville sample (p = 0.016). Right part, multilocus test: the multilocus linkage disequilibrium value is given in the first column, and the statistical probability (p) of no linkage disequilibrium is estimated in the last column.

| Population (n) | # of pairwise LD tests | significant at 1% (expected) | significant at 5% after Bonferroni | Multilocus LD ($Ia^S$) | p value 2000 bootstraps |
|---|---|---|---|---|---|
| Franconville (12) | 105 | 9 (1.0) | 6 | 0.28 | <0.001 |
| Hermanville (12) | 120 | 7 (1.2) | 3 | 0.084 | 0.001 |
| Merlet1+2+3 (19) | 253 | 77 (2.5) | 19 | 0.36 | <0.001 |
| Each multilocus genotype (24) | 666 | 61 (6.7) | 0 | 0.049 | <0.001 |
| Idem w/o singletons (17) | 351 | 48 (3.5) | 14 | 0.065 | 0.001 |
| Idem w/o local polym. (15) | 66 | 7 (0.7) | 5 | 0.070 | <0.001 |

**Table 4. Outcrossing rate in *C. elegans* wild populations.** Each individual is characterized by its microsatellite repeat number (italics) at two loci located on chromosomes II and IV (II-R and IV-R; [15]). Heterozygotes were found in the Franconville and Sainte-Barbe populations. Genotype counts are given in Table S3. Population: location and sampling date (cf. Table 1B). N: number of genotyped individuals; *He*: gene diversity; *Fis*: inbreeding coefficient; *S*: selfing rate. The mean is the mean of the populations.

| Population | N | Allele frequencies | | *He* | *Fis* | *S* |
|---|---|---|---|---|---|---|
| | | *II – R* | *IV - R* | | | |
| Le Perreux 5 Oct 2004 | 41 | *28*: 1 | *69*: 0.610<br>*71*: 0.390 | 0.677 | 1 | 1 |
| Franconville 6 Oct 2004 | 129 | *24*: 0.338<br>*28*: 0.663 | *64*: 0.008<br>*66*: 0.224<br>*68*: 0.08<br>*69*: 0.728<br>*70*: 0.033 | 0.639 | 0.963 | 0.981 |
| Ste Barbe 3 Oct 2004 | 88 | *28*: 0.073<br>*36*: 0.913<br>*43*: 0.013 | *68*: 0.083<br>*69*: 0.917 | 0.415 | 0.958 | 0.979 |
| Primel 3 Oct 2004 | 50 | *24*: 0.023<br>*28*: 0.023<br>*36*: 0.163<br>*37*: 0.023<br>*43*: 0.767 | *60*: 0.021<br>*68*: 0.574<br>*69*: 0.404 | 0.786 | 1 | 1 |
| mean | | | | | 0.974 | 0.987 |

**Supplemental tables**

| Population | Genotype *II-R* | Genotype *IV-R* | # animals | total |
|---|---|---|---|---|
| Le Perreux 5 Oct 2004 | *28/28* | *69/69* | 4 | 41 |
| | *28/28* | *71/71* | 13 | |
| | - | *69/69* | 21 | |
| | - | *71/71* | 3 | |
| Franconville 6 Oct 2004 (compost) | *24/24* | *64/64* | 1 | 129 |
| | *24/24* | *66/66* | 24 | |
| | ***24/24*** | ***66/69*** | **3** | |
| | *24/24* | *68/68* | 1 | |
| | *24/24* | *69/69* | 8 | |
| | *24/24* | - | 3 | |
| | *28/28* | *66/66* | 1 | |
| | *28/28* | *69/69* | 72 | |
| | *28/28* | *70/70* | 3 | |
| | *28/28* | - | 3 | |
| | - | *69/69* | 8 | |
| | - | *70/70* | 1 | |
| | ***24/28*** | ***66/66*** | **1** | |
| Ste Barbe 3 Oct 2004 | *28/28* | *68/68* | 3 | 88 |
| | *28/28* | - | 2 | |
| | *36/36* | *69/69* | 66 | |
| | *36/36* | - | 2 | |
| | *43/43* | *68/68* | 1 | |
| | ***28/36*** | ***69/69*** | **1** | |
| | - | *68/68* | 3 | |
| | - | *69/69* | 10 | |
| Primel 3 Oct 2004 (compost + 7 isopods) | *24/24* | *60/60* | 1 | 50 |
| | *28/28* | *69/69* | 1 | |
| | *36/36* | *68/68* | 3 | |
| | *36/36* | *69/69* | 4 | |
| | *37/37* | *69/69* | 1 | |
| | *43/43* | *68/68* | 19 | |
| | *43/43* | *69/69* | 11 | |
| | *43/43* | - | 3 | |
| | - | *68/68* | 5 | |
| | - | *69/69* | 2 | |